Limitations of zT as a Figure of Merit for Nanostructured Thermoelectric Materials

Xufeng Wang and Mark Lundstrom
Electrical and Computer Engineering
Purdue University
West Lafayette, IN USA 47907

**Abstract** — Thermoelectric properties of nanocomposites are numerically studied as a function of average grain size or nanoparticle density by simulating the measurements as they would be done experimentally. In accordance with previous theoretical and experimental results, we find that the Seebeck coefficient, power factor and figure of merit, *zT*, can be increased by nanostructuring when energy barriers exist around the grain boundaries or embedded nanoparticles. When we simulate the performance of a thermoelectric cooler with the same material, however, we find that the maximum temperature difference, $\Delta T_{\max}$, is much less than expected from the given *zT*. This occurs because the measurements are done in a way that minimizes Joule heating, but the Joule heating that occurs in operating devices has a large effect for these kinds of materials. The same nanocomposite but without energy barriers at the grain boundaries has a lower measured *zT* but a higher $\Delta T_{\max}$. The physical reason for these results is explained. The results illustrate the limitations of *zT* as a figure of merit for nanocomposites with electrically active grain boundaries.

## I. Introduction

To address the challenge of increasing the thermoelectric materials figure of merit, *zT*, Hicks and Dresselhaus suggested in 1993 that *zT* might be enhanced in nanostructures [1, 2]. Following this strategy, researchers have been able to substantially and steadily increase *zT* [3]. Today, there are many reports of $zT > 1$, and all of them make use of nanostructuring of one kind or another [4]. This progress has primarily been achieved by nanostructuring materials to reduce the lattice thermal conductivity without substantially degrading the electrical conductivity [5-8], but benefits to the electronic performance have also been reported. These benefits include an increased Seebeck coefficient (e.g. [9-21] ) and/or a reduction of bipolar effects [6]. Enhanced Seebeck coefficients are thought to be due to energy filtering near nanoscale inclusions or grain boundaries [22]; theoretical treatments [23] and models [24] have been reported. If nanostructured materials could be engineered to both lower the lattice thermal conductivity and improve the electronic performance, then substantial additional increases in *zT* would be possible [4].

The physics of interfaces due to nanoscale inclusions or grain boundaries has been studied and various models have been proposed. In [25], the authors modeled the grain boundary as a parabolic band material with potential barriers for carriers and found that the model could readily explain experimental results. More detailed models have also been reported [26-30]. In [31, 32], the authors showed that recent experimental results in rather complicated thermoelectric materials could be explained using simple models without considering the energy-dependence of the carrier scattering. While these models differ in detail, they all show band bending at interfaces is the origin of differences observed in the Seebeck coefficients of bulk crystals and nanocomposites.

Based on these prior studies, we model the interfaces as electrostatic potential barriers using full, numerical simulations of random structures. The details are discussed in Section II.

Two different types of nanocomposite thermoelectric materials are examined. The first is p-type polycrystalline $Bi_2Te_3$, and the second is p-type crystalline $Bi_2Te_3$ with embedded silver nanoparticles. We use these as model systems for which experimental data is available [17, 33] but the conclusions should be more general. The thermoelectric properties as a function of average grain size or nanoparticle density are examined by numerically simulating the measurements of electrical and thermal conductivity and Seebeck coefficient as they would be done experimentally. In addition, we simulate the performance of an ideal thermoelectric cooler (i.e. no interface resistances or shunt conductances) and compare the $zT$ extracted from $\Delta T_{\max}$ to the $zT$ obtained from the simulated measurements of the thermoelectric parameters.

We focus in this paper on polycrystalline nanocomposites, but similar results (discussed in the supplementary information) are observed for nanocomposites with embedded Ag nanoparticles. For the case where the grain boundary acts to deplete the grain, energy barriers for majority carriers result, and we find in accordance with experiments, that the Seebeck coefficient, power factor and figure of merit, $zT$, can be increased by nanostructuring, but when we simulate the performance of a thermoelectric cooler, we find that the maximum temperature difference, $\Delta T_{\max}$, is much less that what would be expected from the given $zT$. An important finding is that conventional measurements of the thermoelectric parameters of nanostructured thermoelectric materials may provide overly optimistic predictions of device performance. This occurs when the grain boundaries are electrically active because Joule heating, which is negligible in the $zT$ measurement plays a strong role under device operating conditions. The same nanocomposite but with no potential barriers at the grains has a lower $zT$ but produces a higher $\Delta T_{\max}$ in a TE cooler. The physical reason for these results will be explained.

The paper is organized as follows. Section II describes the numerical techniques for the two-dimensional electrothermal simulations. Also described in Sec. II are the material parameters used in the simulations and the treatment of grain boundaries. Section III presents the results in terms of $zT$ vs. grain size where $zT$ is extracted three ways: 1) by simulating the measurement of the individual thermoelectric parameters, 2) by simulating Harman method measurements [34], and 3) by deducing $zT$ from the $\Delta T_{\max}$ of a simulated cooler. In Sec. IV we discuss the results and summarize the conclusions in Sec. V.

## II. Approach

In this study, we use the semiconductor device simulation program, Sentaurus, which solves the coupled partial differential equations that describe electrothermal transport in semiconductors [35]. The following equations [36] were numerically solved:

$$\nabla \bullet \vec{D} = \rho \quad \text{(1a)}$$

$$\frac{\partial n}{\partial t} = -\nabla \bullet \left( \vec{J}_n / -q \right) - R \quad \text{(1b)}$$

$$\frac{\partial p}{\partial t} = -\nabla \bullet \left( \vec{J}_p / \; q \right) - R \quad \text{(1c)}$$

$$c_{tot} \frac{\partial T}{\partial t} = \nabla \cdot \left( \kappa_{tot} \vec{\nabla} T \right) + H. \quad \text{(1d)}$$

Equation (1a) is Poisson's equation, where $\vec{D}$ is the displacement vector and $\rho$ the space charge density, which includes contributions from mobile electrons and holes and ionized dopants. Equations (1b) and (1c) are the electron and hole continuity equations where $\vec{J}_n$ and $\vec{J}_p$ are the current densities that include contributions from gradients in the electrochemical potential and temperature ($q$ is the magnitude of the elementary charge). The term, $R$, is the carrier recombination rate, which is only important in the presence of bipolar effects. The final equation, (1d), is an energy balance equation for the temperature, $T$. Local thermodynamic equilibrium is assumed so that the electron, hole, and lattice temperatures are identical. The total heat capacity, $c_{tot}$, and the total thermal conductivity, $\kappa_{tot}$, include contributions from the lattice as well as the electrons and holes. Finally, the heat source term, $H$, includes contributions from Joule heat, recombination heat, and Thompson heat [36]. Transient simulations are performed only to simulate the Harman method – all other simulations are steady-state.

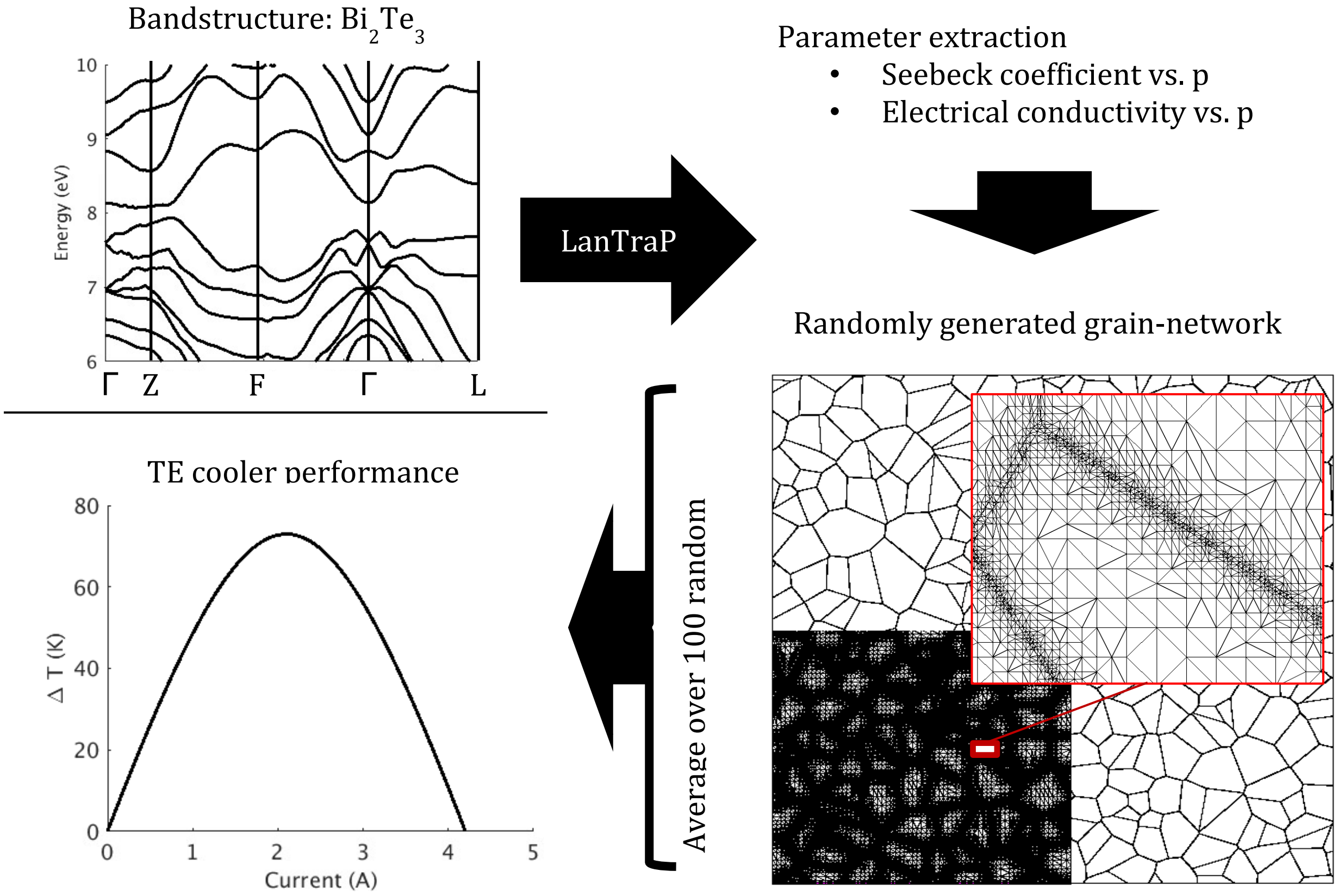

Fig. 1. Simulation flow used in this work. Upper left: Bandstructure as obtained from DFT simulation. Upper right: Parameter extraction using the LanTraP tool [9]. Lower right: Grain structure and numerical grid for one randomly generated sample. The dark areas show the high density numerical grid. The inset shows an expanded view of how the grid is refined near and inside the grain boundaries. The average grain size here is 1 micrometer, and the simulation domain is 1 mm by 1 mm. Lower left: Simulated performance of a TE cooler. For a typical case, 100 random samples were generated and simulated and the results were averaged.

**Material-level simulation and material parameters**

In this work, $Bi_2Te_3$ was chosen as a model thermoelectric material, because it has a simple rhombohedral crystal structure, its material properties are relatively well-known [37], and experimental results are available [17, 34]. The methods and results obtained in this work, should, however, be applicable to other thermoelectric materials as well. The overall simulation workflow is summarized in Fig. 1.

Because thermoelectric materials have highly non-parabolic and complex band structures, the first step was to use density functional theory (DFT) calculations with Quantum Espresso [38] to obtain the $Bi_2Te_3$ band structure. For both Bi and Te, we used full relativistic PAW pseudo-potentials with a 40 Ry plane wave energy cutoff and spin-orbit coupling. An 8×8×8 and 20×20×20 Monkhorst-Pack k-point mesh was used for self-consistent and non-self-consistent calculations respectively.

The second step was to process the full bandstructure with LanTraP, a program that extracts thermoelectric transport parameters from a given band structure [9]. We assumed that the scattering rate for carriers was proportional to the density-of-states and determined the electron-phonon coupling parameter by matching the resistivity to experimental values [39]. Transport parameters were extracted along the cross-plane direction. We did not include the transport anisotropy of $Bi_2Te_3$, because our focus was on the effects of grain boundaries, and the random structures should average out anisotropies.

The simulated electrical resistivity and Seebeck coefficient vs. hole density of p-type, crystalline $Bi_2Te_3$ are shown in Fig. 2. These parameters are similar to reported values [40-42]. The maximum power factor (PF) occurs at a hole concentration of $4\times10^{19}$ cm$^{-3}$. For the subsequent simulations of polycrystalline materials, each grain was doped at this optimal point to ensure that any performance increase observed in the polycrystalline material was due to grain boundary effects.

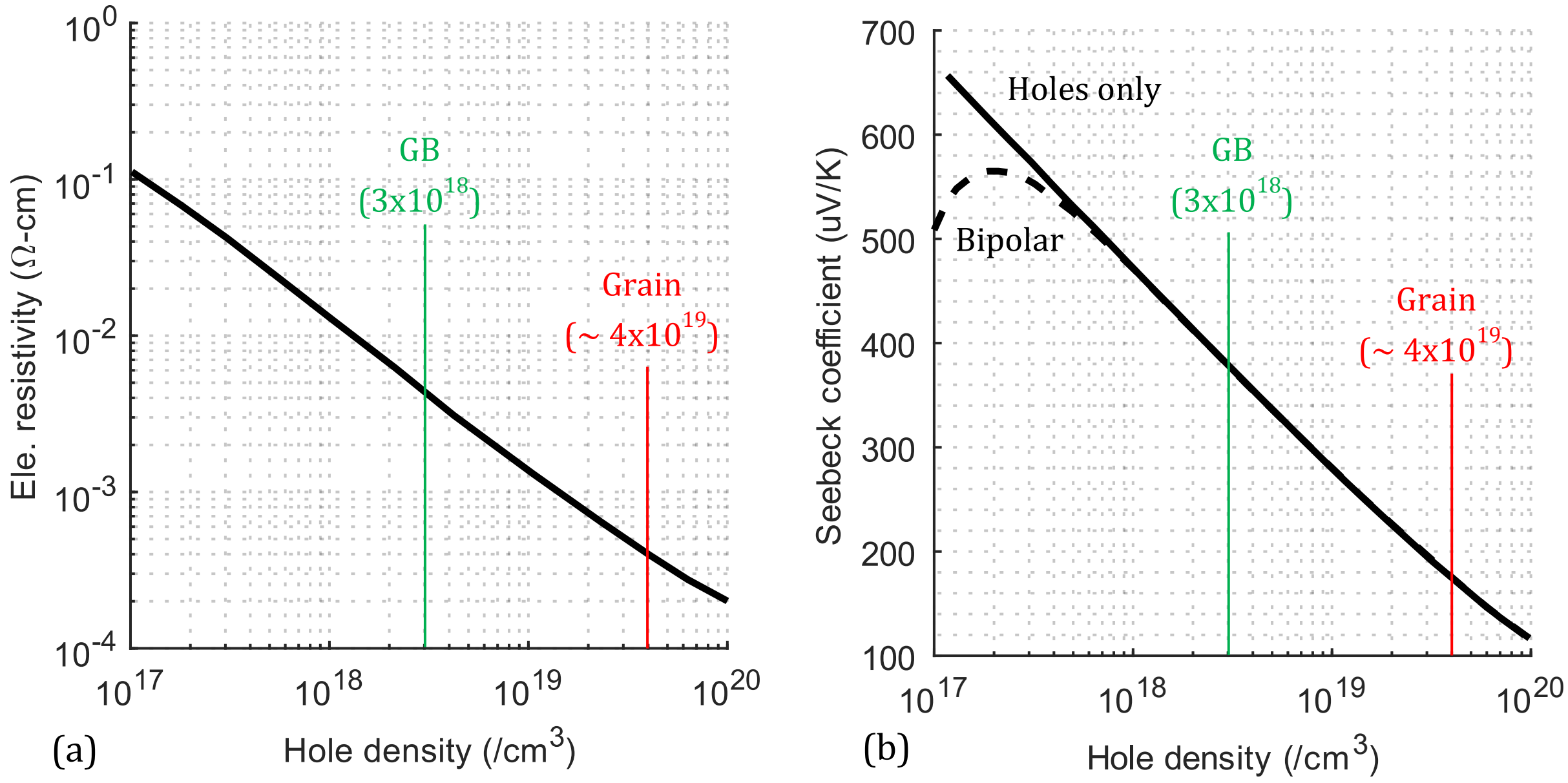

Fig. 2. (a) Electrical resistivity and (b) Seebeck coefficient vs. hole concentration for the $Bi_2Te_3$ baseline material used in this work. The bulk is doped at the optimal PF point, $N_A^G = 4\times10^{19}$ cm$^{-3}$. To produce depleted grain boundary regions, $N_A^{GB} = 3\times10^{18}$ cm$^{-3}$ was used.

**Electrical simulation of nanocomposite materials**

To simulate nanocomposites, a random distribution of grains was first generated, and then the grain boundaries were defined. The algorithm used to generate this random geometry is described in the supplementary material. The grain boundaries were 100 nm thick. A simple thermal interface resistance was added in the middle of each grain boundary region to account for the additional thermal resistance. The value of this thermal interface resistance was set to agree with experimentally reported trends in thermal resistivity vs. grain density [17, 43]. The doping density of the grain boundaries was set at $N_A^{GB} = 3\times10^{18}$ cm$^{-3}$ to produce results consistent with experiments. The resulting band bending is about one-half of the band gap. As shown in Fig. 2(b), the result was that the Seebeck coefficient of the grain boundaries was substantially higher than that of the grains.

The generated polycrystalline geometry was used as an input to the Sentaurus Device simulator. It was critical to resolve the band bending within and near the grain boundaries. For polycrystalline materials, a fine mesh on both sides of the grain boundary was needed. The added burden of a dense mesh is the main reason why we used 2D simulations instead of 3D. A full 3D simulation of a geometry that is large enough to be statistically meaningful would have taken days. As discussed in Sec. VI, simulations of 1D nanocomposites show results that are very similar to those obtained by 2D simulations; it is likely, therefore, that the effects found in the 2D studies would not change in 3D. (See the supplementary information for a description of how the 1D simulations were done.)

The Sentaurus simulations used tables of transport parameters vs. carrier density for all thermoelectric parameters; these tables were generated using LanTraP with the aforementioned DFT calculations. Phenomena such as the Thomson effect and mobility variations with doping density were, therefore, automatically included. For each random sample, we computed the conductivity by applying a small voltage (0.1 mV), computing the resulting current, and converting the result to a conductivity for the sample. To compute the Seebeck coefficient and thermal conductivity, we applied a small temperature difference (1 K), and simulated the resulting open circuit voltage, which yields the Seebeck coefficient and heat flux, from which, we determine the thermal conductivity. From the three parameters, we could compute a material figure of merit, *zT*, for the sample. We also determined the *zT* by simulating a Harman measurement [34] for each sample. Finally, we extracted a *zT* by simulating the $\Delta T_{\max}$ of an ideal 1-leg TE cooler. The results of 100 such simulations for each average grain size were averaged and plotted.

## III. Results

The simulated Seebeck coefficient for polycrystalline $Bi_2Te_3$ as a function of average grain size is shown in Fig. 3(a). Because the simulation domain (1 mm by 1 mm) is large enough to contain thousands of grains, significant grain size averaging occurs for each randomly generated sample. Still, for each average grain size, 100 randomly generated samples were simulated and averaged. The error bars shown in Fig. 3(a) show that the sample-to-sample variation is small. As intuitively expected, with decreasing average grain size, the Seebeck coefficient increases because the grain boundary regions have higher Seebeck coefficients, and as the average grain size decreases, there are more grain boundaries. Still, the size of the grain boundaries is small, so the grain boundary regions occupy a small fraction of the polycrystalline sample. If we repeat the simulation but assume no reduction in lattice thermal conductivity, then as Fig. 3(a) shows, there is a much smaller dependence of the Seebeck coefficient on grain size.

As noted above, the Seebeck coefficient of the sample increases substantially when the Seebeck coefficient of the grain boundary is high and the thermal conductivity of the grain boundary is low [27, 33]. This can be understood with a simple 1D model [33]:

$$S = \frac{S_G\left(1-t_{GB}\right)/\kappa_G + S_{GB}\left(t_{GB}/\kappa_{GB}\right)}{\left(1-t_{GB}\right)/\kappa_G + \left(t_{GB}/\kappa_{GB}\right)} , \quad (2)$$

where $S_G\left(S_{GB}\right)$ is the Seebeck coefficient of the grain (grain boundary), $t_{GB}$ is the thickness of the grain boundary, and $\kappa_G\left(\kappa_{GB}\right)$ is the thermal conductivity of the grain (grain boundary). Figure 2 shows that $S_{GB} \approx 2S_G$, but the grain boundary is thin, that it has only a small effect on the overall Seebeck coefficient unless the thermal conductivity of the grain is also very low. This expectation is confirmed by the results shown Fig. 3(a) – when the thermal conductivity of the grain boundaries is identical to the thermal conductivity of the grains, the Seebeck coefficients of the polycrystalline samples are close to that of the bulk, crystalline reference.

Figure 3(b) shows the electrical resistivity vs. average grain size. The electrical resistivity is more sensitive to grain size than is the Seebeck coefficient, and it is independent of whether or not the lattice thermal conductivity is decreased. This result can also be understood with a simple, 1D model [33]:

$$\rho = \left(1-t_{GB}\right)\rho_G + t_{GB}\rho_{GB} . \quad (3)$$

Grain boundaries are thin, but as Fig. 2(b) shows, the resistivity of the grain boundaries is more than an order of magnitude larger than that of the grain. This occurs because band bending in the grain boundary exponentially decreases the carrier concentration (it linearly increases the Seebeck coefficient). The smaller the grains are, the more of these grain boundary regions exist, and thus the higher the overall resistivity is. As a result, electrical resistivity is inversely proportional to the average grain size.

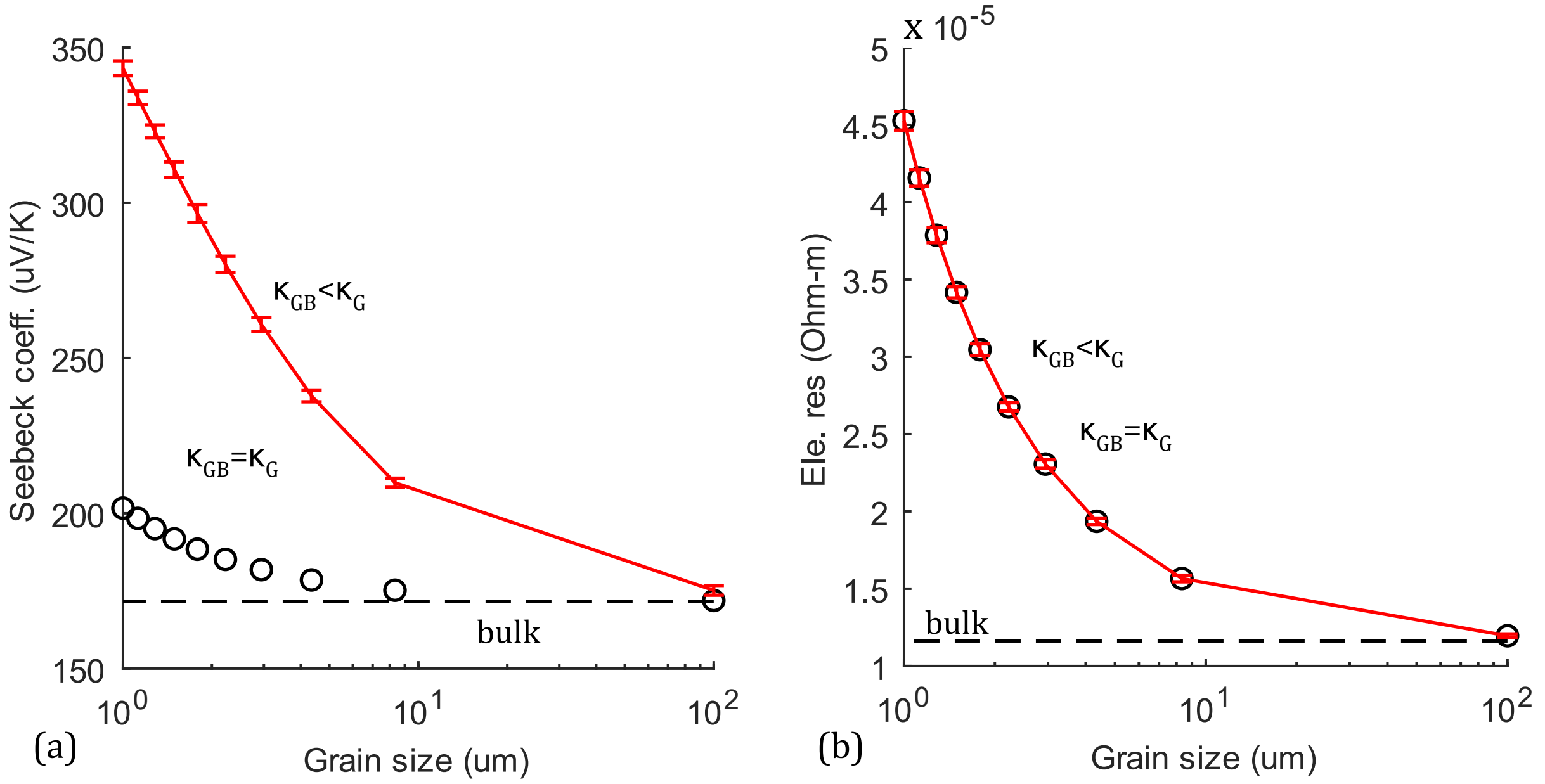

Fig. 3. (a) Seebeck coefficient and (b) Electrical resistivity vs. average grain size for 2D polycrystalline $Bi_2Te_3$. Also shown for reference are the corresponding results for homogenous bulk $Bi_2Te_3$ (dashed line). Finally, results for polycrystalline samples with no reduction of the thermal conductivity are also shown (open circles).

The thermal conductivity shown in Fig. 4(a) can also be explained in the same way: the more grain boundaries there are, the more interface scattering for phonons occurs, and thus the lower the thermal conductivity is. The power factor calculated using the aforementioned Seebeck coefficient, electrical resistivity, and thermal conductivity is shown in Fig. 4(b). For the particular set of input parameters chosen, *PF* exhibits a peak value at an average grain size of 3 micrometers. For smaller grain sizes, the increase in resistivity begins to dominate, and *PF* decreases. This echoes reports in the literature that show an optimal grain sizes exist for various thermoelectric nanocomposites [42-44].

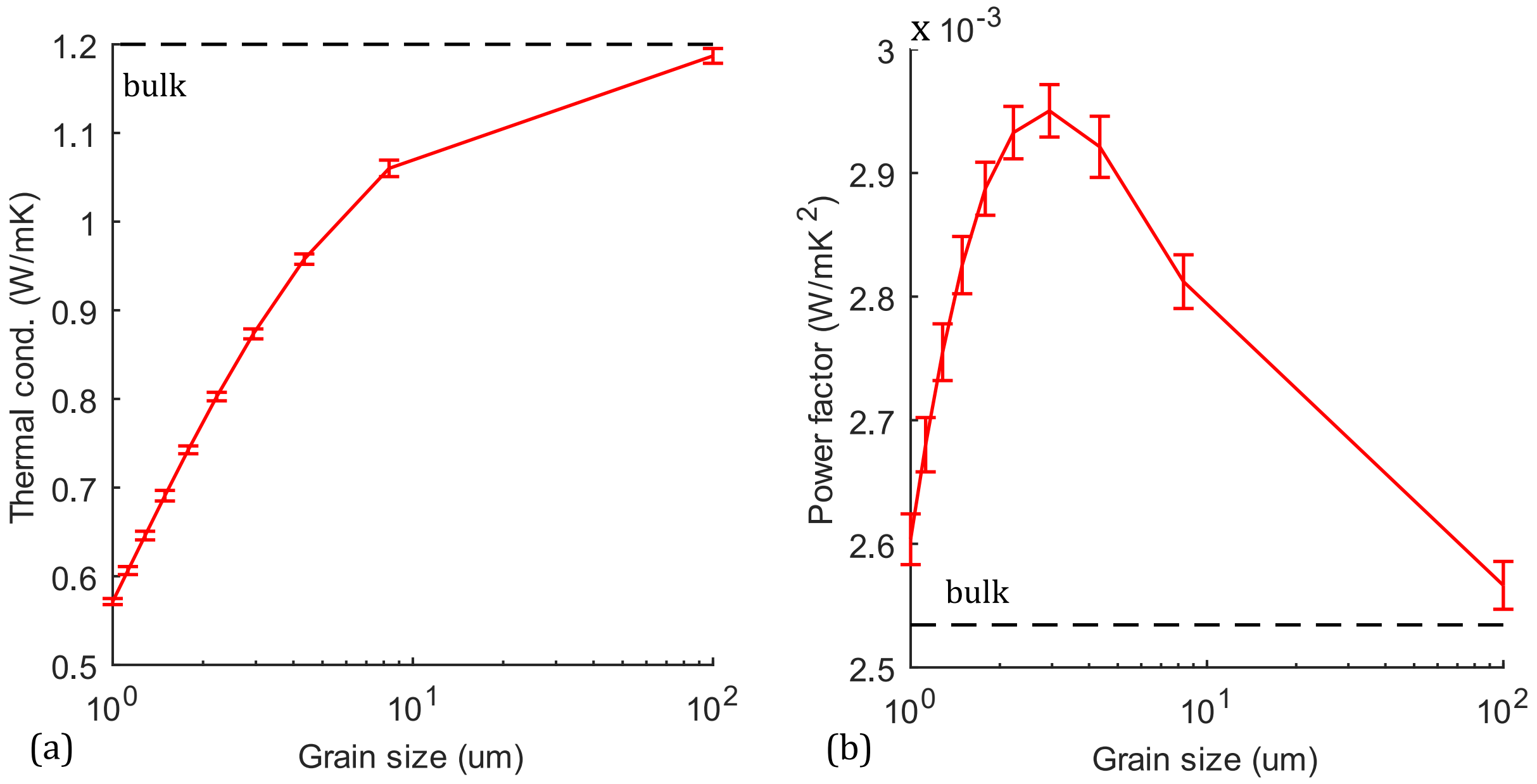

Fig. 4. (a) Thermal conductivity and (b) Power factor vs. average grain size for 2D polycrystalline $Bi_2Te_3$. Also shown for reference are the corresponding results for homogenous bulk $Bi_2Te_3$ (dashed line).

Finally, we examine the thermoelectric figure of merit of the simulated nanocomposites. The figure of merit was obtained in three different ways. The first was from the simulated measurements of the electrical conductivity, $\sigma$, the Seebeck coefficient, $S$, and the total thermal conductivity, $\kappa$ according to the definition:

$$zT(\sigma,S,\kappa)=\frac{\sigma S^2 T}{\kappa}, \tag{4}$$

where $T$ is the temperature. This is the method most commonly used to report the measured $zT$ of nanocomposite materials. The second way $zT$ was determined was from simulated Harman measurements, which we label as $zT(\text{Harman})$. The third way $zT$ was determined was from the maximum cooling temperature, $\Delta T_{max}$, of a simulated thermoelectric cooler,

$$\Delta T_{max}=\frac{zT_C{}^2}{2}. \tag{5}$$

Because the simulated cooler is ideal with no contact resistances or thermal shunts, we expect values for $zT(\Delta T_{\max})$ that are similar to those obtained with the first two methods. (A small difference might be expected because of the temperature dependence of the thermoelectric parameters.) The key question in this paper is whether the $zT(\sigma,S,\kappa)$ predicted from Eq. (4), $zT(\text{Harman})$, and $zT(\Delta T_{\max})$ deduced from eqn. (5) are the same. Figure 5 provides the answer.

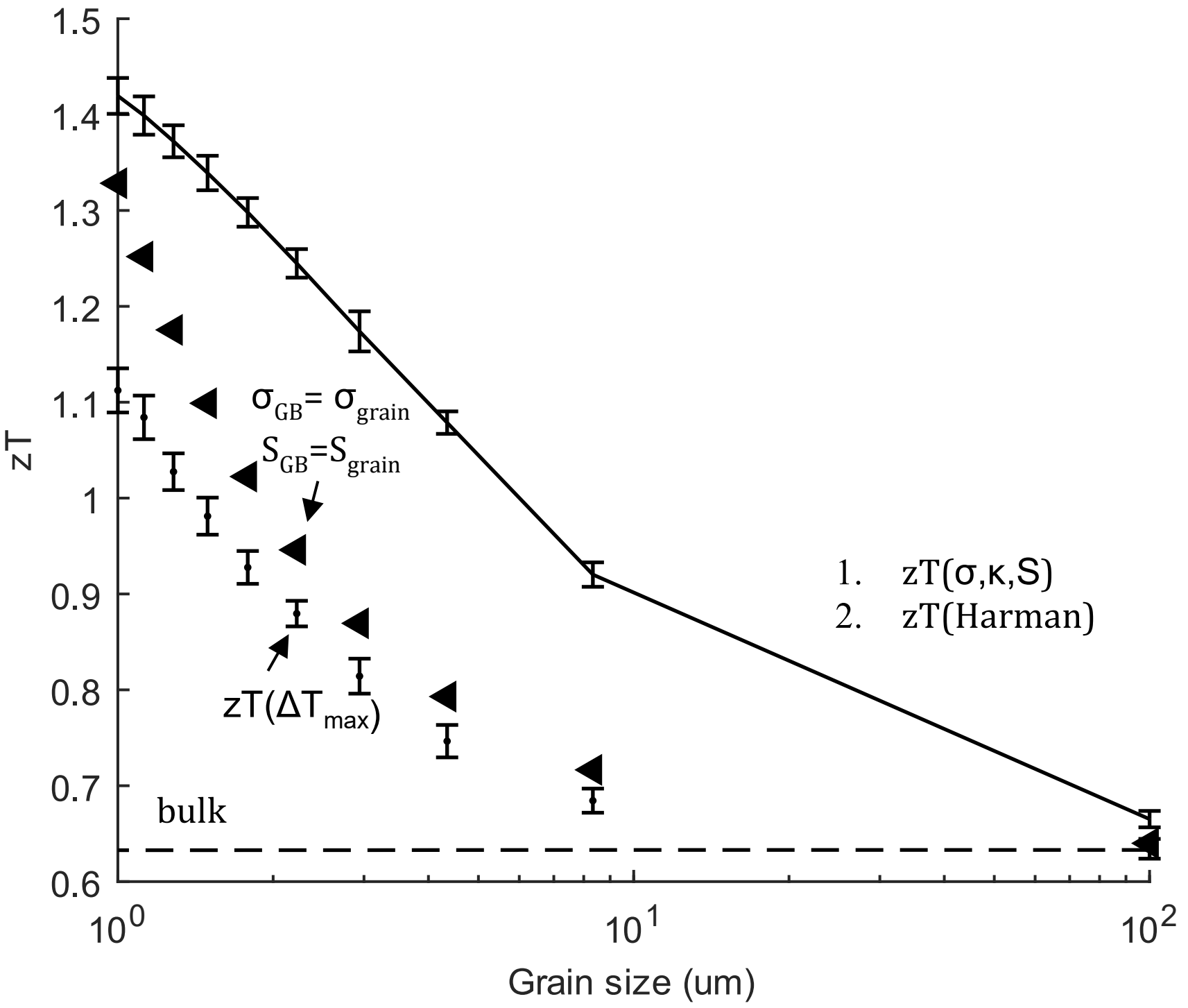

Fig. 5. Figure of merit, *zT,* vs. average grain size for nanocomposite $Bi_2Te_3$. Also shown for reference are the corresponding results for homogenous bulk $Bi_2Te_3$ (dashed line). The results for $zT(\sigma, S, \kappa)$ and $zT(\text{Harman})$ are substantially larger than for $zT(\Delta T_{\text{max}})$, but all three results for the nanocomposites are larger than the corresponding results for homogenous bulk $Bi_2Te_3$. Results for polycrystalline samples for which the electrical properties of the grains and grain boundaries are identical are shown as closed triangles. In this case, $zT(\sigma, S, \kappa)$, $zT(\text{Harman})$, and $zT(\Delta T_{\text{max}})$ are identical.

As shown in Fig. 5, the $zT(\sigma, S, \kappa)$ and $zT(\text{Harman})$ are identical but significantly larger than $zT(\Delta T_{\text{max}})$ for all average grain sizes. In other words, the $\Delta T_{max}$ of the cooler built from polycrystalline material would not be as high as would be expected As discussed in the supplementary information, the conclusion for $Bi_2Te_3$/Silver composites is the same.

Also shown in Fig. 5 are results for the case in which the electrical properties of the grain boundaries are identical to those of the grains and the crystalline reference. For this case, the only difference between the polycrystalline and crystalline samples is the lattice thermal conductivity. Figure 5 shows that for this case, *zT* is greater than for the crystalline reference and that all three methods of obtaining *zT* give identical results.

## VI. Discussion

The difference between the $zT$ predicted from the individually measured electrical and thermal conductivities and Seebeck coefficient (or from a Harman measurement) and the $zT$ deduced from the cooler operating at $\Delta T_{\max}$ originates from the nature of the measurements. Seebeck coefficients are measured by creating a temperature gradient across a sample and then measuring the open-circuit voltage. To minimize the effects of temperature dependent material parameters, the temperature gradient applied across the sample is typically very small – a few Kelvin [45, 46]. Harman measurements are typically done at low currents to minimize Joule heating. A thermoelectric cooler operates at high currents where Joule heating is significant.

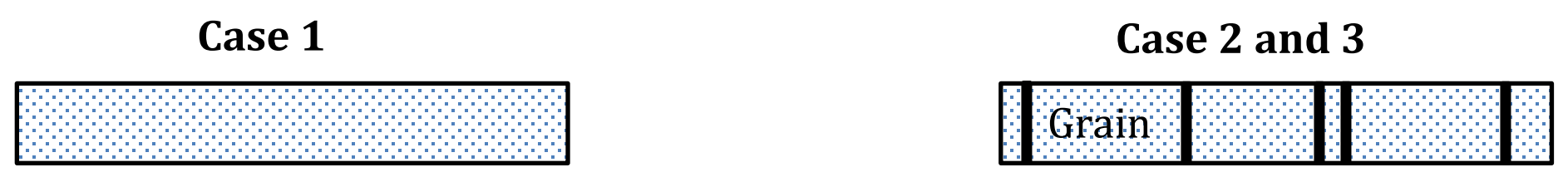

Fig. 6. Illustration of 1D nanocomposites. The average grain size is 1.3 $\mu$m, and the material properties of the grains and grain boundaries are the same as for the 2D samples considered in Fig. 5. Case 1: uniform material. Case 2: nanocomposite with reduced $\kappa$ in grain boundaries. Case 3: nanocomposite with reduced $\kappa$ and enhanced $S$ in grain boundaries.

One-dimensional simulations of the structure shown in Fig. 6 provide some insights into these results. Three different cases were considered: 1) A uniform material with the properties of the grains in the 2D simulations, 2) a 1D nanocomposite with reduced thermal conductivity in the grain boundaries, but with electrical properties that are idential to those of the grain, and 3) a 1D nanocomposite with a reduced thermal conductivity in the grain boundaries and with the lighter doping that produces potential barriers for carriers.. Case 1) is the bulk reference. Case 2) is a nanocomposite that benefits only from a reduction of the lattice thermal conductivity, and Case 3) is a nanocomposite that also provides an enhanced power factor. For case 3, the 1D simulations give $zT(\sigma, S, \kappa) = zT(\text{Harman}) = 1.35$ and $zT(\Delta T_{\max}) = 1.02$, which, for an average grain size of 1.3 $\mu$m, agrees well with the 2D results shown in Fig. 5.

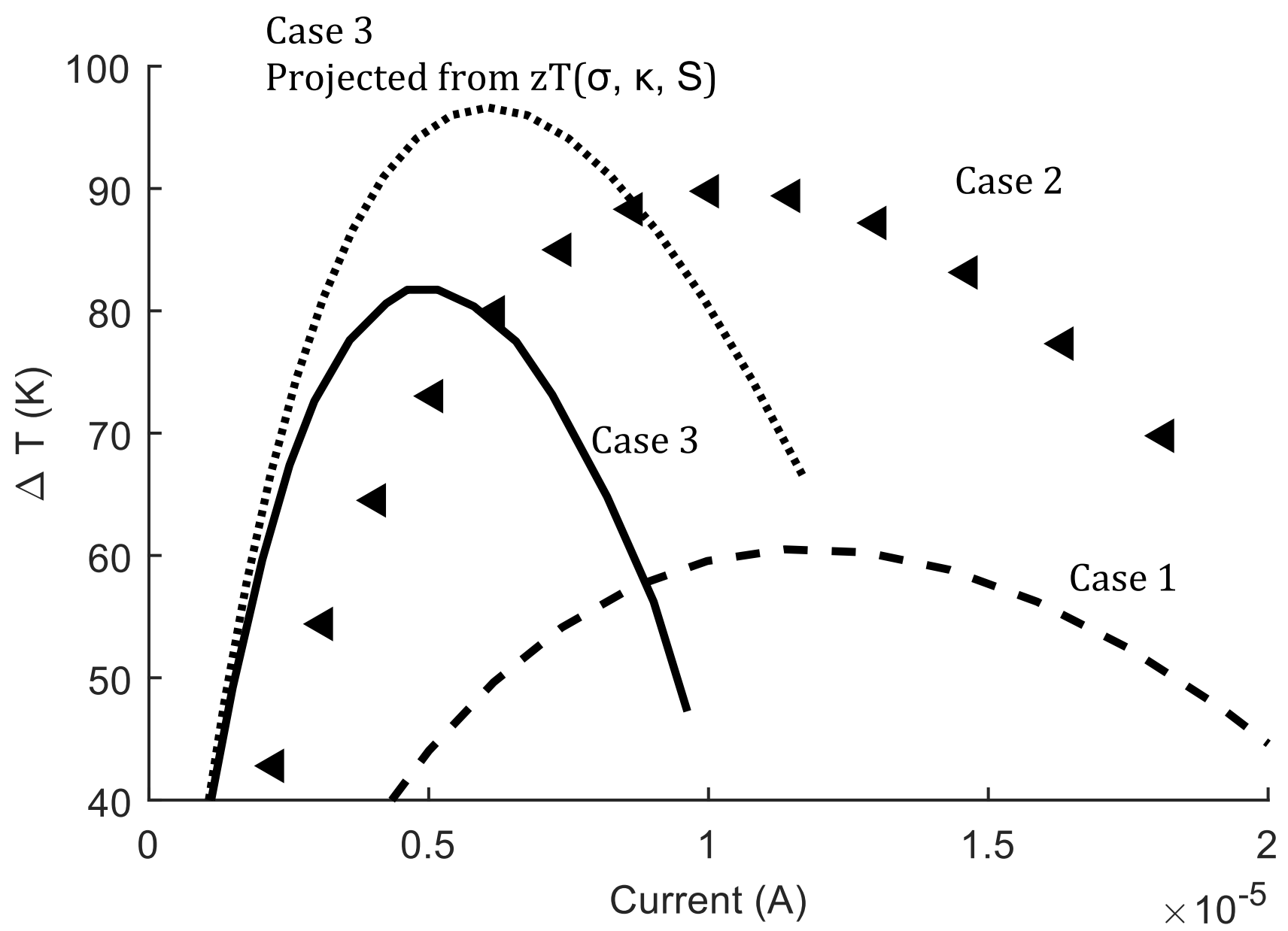

Fig. 7. Simulated $\Delta T$ *vs.* $I$ for a thermoelectric cooler with the three cases shown in Fig. 6. Also shown for reference is the projected $\Delta T$ *vs.* $I$ assuming $zT(\sigma, S, \kappa) = 1.35$, which conventional measurements would give for the Case 3 nanocomposite.

Figure 7 shows the $\Delta T$ *vs.* $I$ comparison for the three cases shown in Fig. 6. As expected, Case 1, the bulk crystal case, has the lowest performance. For Case 2, when the grains only have the added benefits of increased thermal resistance while their electrical properties are identical to those of grains, the cooling performance is significantly enhanced. Finally, note that Case 3, for which the grains have an enhanced $S$ as well as a lower thermal conductivity, gives a lower performance than Case 2. This occurs in spite of the fact that the conventionally measured *zT* for Case 3 is $zT(\sigma, S, \kappa) = 1.35$ while that for Case 2 is $zT(\sigma, S, \kappa) = 1.05$. Fig 7 also shows the $\Delta T$ *vs.* $I$ that would be predicted from the conventionally measured *zT* (1.35) for the Case 3 nanocomposite. These results demonstrate that conventional measurements of nanocomposites with electrically active grain boundaries (Case 3) substantially overpredict the performance of thermoelectric coolers (compare the solid and dashed lines in Fig. 7). They also show that while potential barriers at grain boundaries can increase the measured *zT*, they decrease the performance of a thermoelectric cooler in comparison to a nanocomposite with no potential barriers at the grain boundaries (i.e. Case 3 vs. Case 2).

One way to understand these effects is in terms of the compatibility factor discussed by Snyder, *et al.* [47]. The grains are optimally doped to have the highest power factor according to Fig. 2, but the grain boundaries are not optimally doped. Intuitively one might expect this effect to be small because the grain boundaries occupy a small portion of the overall geometry. The difference shown in Fig. 7 between Case 2 and Case 3 is, however, not insignificant. The additional performance degradation can be understood as due to the incompatibility between the grain and grain boundary elements. The essential point-of-view of the compatibility factor is to analyze the 1D thermoelectric device as infinitesimally small segments cascaded in series. Next, we adopt this view and compare the thermoelectric device under Seebeck measurement conditions and at the $\Delta T_{max}$ of a thermoelectric cooler.

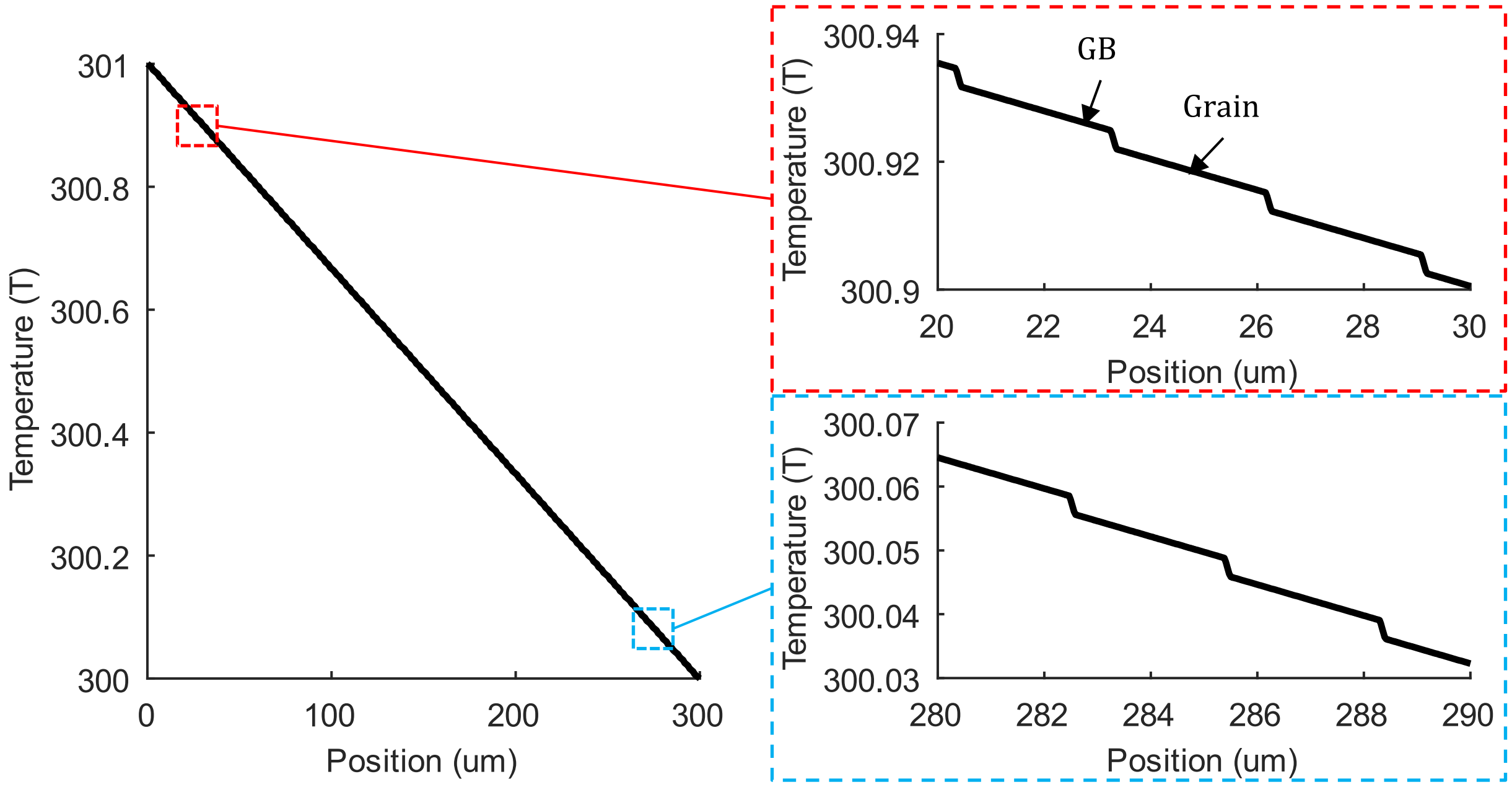

Fig. 8. Temperature vs. position for the Case 3) nanocomposite under open-circuit conditions with $\Delta T = 1$ K. These are the conditions used for Seebeck coefficient measurements.

As shown in Fig. 8 for the Seebeck measurement, the thermoelectric device is subject to a very small temperature difference, and the device is open-circuited electrically. The temperature profile shows the expected linear profile. The details of grain boundaries are visible in the insets of Fig. 8. Since grain boundary regions have high thermal resistance, more temperature drops across them, and the profile shows a zig-zag pattern. As discussed earlier in Eq. (2), this causes the overall Seebeck measurement to favor the Seebeck value of the grain boundaries, which is higher than that of the grains.

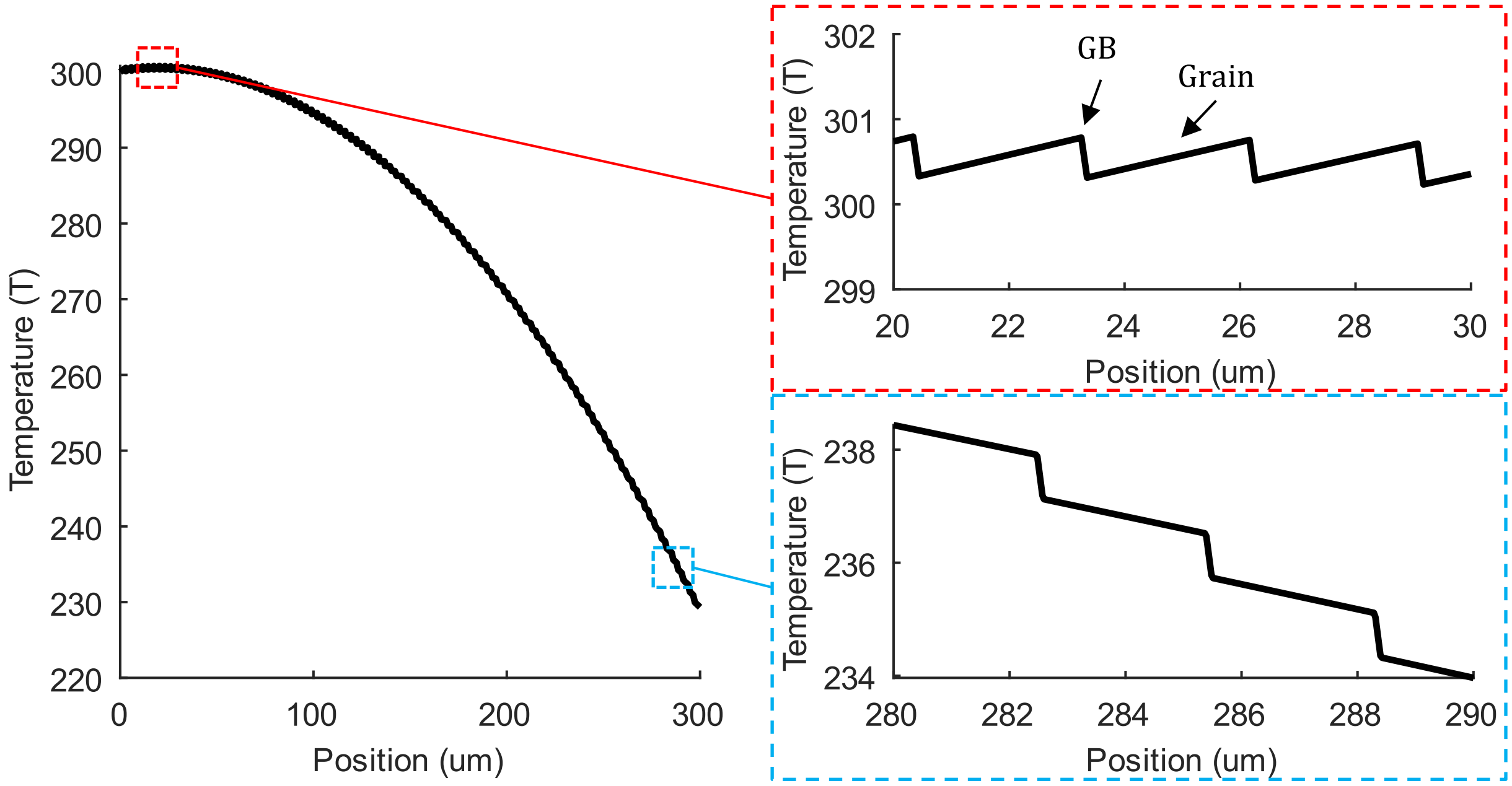

Fig. 9. Temperature vs. position for the Case 3) nanocomposite operating as a thermoelectric cooler with $\Delta T_{\max} = 71$ K. These are the conditions used for Seebeck coefficient measurements.

Under cooling operation, the electrical current is high, and the temperature variations are steeper than in the Seebeck measurement. There is also a significant amount of Joule heating, which cannot be ignored. The temperature profile at $\Delta T_{\max}$ is shown in Fig. 9. Overall, the profile behaves as expected for a homogenous device. The details, however, are important. At the cold side, the profile looks like the one seen for the Seebeck measurement, albeit with a higher temperature variation. Both the grain and grain boundary regions operate as coolers, and the cooling adds up. At the hot side, the profile shows a sawtooth pattern. More importantly, the grain regions no longer act as coolers but as heaters instead. This occurs because the high resistance of the grain boundaries makes them significant Joule heating sources and the low thermal conductance makes it hard for the heat to flow away. In other words, the overall cooling device consists of many small segments of coolers in series with several heaters. This, of course, lowers the cooling performance.

To understand quantitatively why the temperature profile shown in Fig. 8 enhances the sample Seebeck coefficient and why the temperature profile in Fig. 9 does not, consider a one-dimensional example for which the hole current density is

$$J_p = \sigma_p \mathcal{E} - \sigma_p S_p \, dT/dx \,. \tag{4}$$

From the open-circuit voltage for a sample of length, $L$, we can deduce the Seebeck coefficient for the polycrystalline sample from

$$S_{\text{sample}} = \frac{V_{OC}}{\Delta T} = \frac{-\int_0^L S_p \left( dT/dx \right) dx}{\Delta T} \,. \tag{5}$$

Equation (5) shows that the Seebeck coefficient of the sample is strongly influenced by the Seebeck coefficient of regions where the temperature gradient large. When the Seebeck coefficient is measured under open-circuit conditions (Fig. 8), the temperature gradient in the grain boundaries is larger than in the grains, so the overall $S$ increases. On the other hand, at $\Delta T_{\max}$ for the TEC, Joule heating produces a temperature gradient that changes sign in the grain boundary. According to eqn. (5), this change in sign of the temperature gradient essentially eliminates the higher Seebeck coefficient of the grain boundary.

## V. Conclusions

In this work, we examined the electronic performance of nanostructured thermoelectric materials using first-principles informed two-dimensional numerical simulation. It is now well-established that phonon scattering from grain boundaries or imbedded nanoparticles lowers the lattice thermal conductivity, which increases the material figure of merit, *zT*. Our focus was on the electronic performance of these materials – specifically on materials in which the depletion of carriers near grain boundaries (or around embedded nanoparticles) increases the Seebeck coefficient. Simulations of these materials with electrically active grain boundaries confirm that when the Seebeck coefficient of the grain boundaries is larger than that of the grains, the overall Seebeck coefficient of the sample only increases significantly when the lattice thermal conductivity is low where the Seebeck coefficient is high. More importantly, we showed that when the *zT* of the nanocomposite is determined from standard measurements of $\sigma$, $\kappa$, and $S$, it provides an overly optimistic estimate the performance of a thermoelectric cooler. When we compare a nanocomposite with reduced lattice thermal conductivity due to grain boundary scattering to another nanocomposite that also has potential barriers at the grain boundaries, we found that the one with potential barriers had a higher measured *zT*, but it produced a lower $\Delta T_{\max}$ in a TE cooler.

The parameter space for thermoelectric nanocomposites is large. We have explored only the case for which the grains are optimally doped and the grain boundaries deplete the grains and bend the bands by about one-half of the bandgap. Nevertheless, we expect the conclusions to apply more broadly – potential barriers increase the measured *zT* of a nanocomposite but reduce the performance of a thermoelectric cooler made from the same nanocomposite. Electrically active grain boundaries of the type examined here are particularly sensitive to Joule heating, which occurs during device operation but not during the measurement of *zT*. Although we only examined thermoelectric coolers, the same conclusion may also apply to electrical power generation from thermoelectric devices.

Acknowledgements – Authors would like to acknowledge the insightful discussions with Ali Shakouri and Kazuaki Yazawa.

**References**

[1] L. D. Hicks and M. S. Dresselhaus, "Effect of quantum-well structures on the thermoelectric figure of merit," *Physical Review B,* vol. 47, pp. 12727-12731, 1993.

[2] L. D. Hicks and M. S. Dresselhaus, "Thermoelectric figure of merit of a one-dimensional conductor," *Physical Review B,* vol. 47, pp. 16631-16634, 1993.

[3] J. P. Heremans, M. S. Dresselhaus, L. E. Bell, and D. T. Morelli, "When thermoelectrics reached the nanoscale," *Nat Nanotechnol,* vol. 8, pp. 471-3, Jul 2013.

[4] M. G. Kanatzidis, "Nanostructured Thermoelectrics: The New Paradigm?†," *Chemistry of Materials,* vol. 22, pp. 648-659, 2010.

[5] W. Kim, J. Zide, A. Gossard, D. Klenov, S. Stemmer, A. Shakouri*, et al.*, "Thermal conductivity reduction and thermoelectric figure of merit increase by embedding nanoparticles in crystalline semiconductors," *Phys Rev Lett,* vol. 96, p. 045901, Feb 3 2006.

[6] B. Poudel, Q. Hao, Y. Ma, Y. Lan, A. Minnich, B. Yu*, et al.*, "High-thermoelectric performance of nanostructured bismuth antimony telluride bulk alloys," *Science,* vol. 320, pp. 634-8, May 2 2008.

[7] Y. Lan, A. J. Minnich, G. Chen, and Z. Ren, "Enhancement of Thermoelectric Figure-of-Merit by a Bulk Nanostructuring Approach," *Advanced Functional Materials,* vol. 20, pp. 357-376, 2010.

[8] K. Biswas, J. He, Q. Zhang, G. Wang, C. Uher, V. P. Dravid*, et al.*, "Strained endotaxial nanostructures with high thermoelectric figure of merit," *Nat Chem,* vol. 3, pp. 160-6, Feb 2011.

[9] M. S. Dresselhaus, G. Chen, M. Y. Tang, R. G. Yang, H. Lee, D. Z. Wang*, et al.*, "New Directions for Low-Dimensional Thermoelectric Materials," *Advanced Materials,* vol. 19, pp. 1043-1053, 2007.

[10] J. Martin, L. Wang, L. Chen, and G. S. Nolas, "Enhanced Seebeck coefficient through energy-barrier scattering in PbTe nanocomposites," *Physical Review B,* vol. 79, 2009.

[11] D. K. Ko, Y. Kang, and C. B. Murray, "Enhanced thermopower via carrier energy filtering in solution-processable Pt-Sb2Te3 nanocomposites," *Nano Lett,* vol. 11, pp. 2841-4, Jul 13 2011.

[12] B. Paul, A. K. V, and P. Banerji, "Embedded Ag-rich nanodots in PbTe: Enhancement of thermoelectric properties through energy filtering of the carriers," *Journal of Applied Physics,* vol. 108, p. 064322, 2010.

[13] D. Zhang, J. Lei, W. Guan, Z. Ma, C. Wang, L. Zhang*, et al.*, "Enhanced thermoelectric performance of BiSbTe alloy: Energy filtering effect of nanoprecipitates and the effect of SiC nanoparticles," *Journal of Alloys and Compounds,* vol. 784, pp. 1276-1283, 2019.

[14] J. Yang, J. Yan, G. Liu, Z. Shi, and G. Qiao, "Improved thermoelectric properties of n-type Bi2S3 via grain boundaries and in-situ nanoprecipitates," *Journal of the European Ceramic Society,* vol. 39, pp. 1214-1221, 2019.

[15] K. H. Lim, K. W. Wong, Y. Liu, Y. Zhang, D. Cadavid, A. Cabot*, et al.*, "Critical role of nanoinclusions in silver selenide nanocomposites as a promising room temperature thermoelectric material," *Journal of Materials Chemistry C,* vol. 7, pp. 2646-2652, 2019.

[16] N. S. Chauhan, S. Bathula, A. Vishwakarma, R. Bhardwaj, B. Gahtori, A. K. Srivastava, *et al.*, "A nanocomposite approach for enhancement of thermoelectric performance in Hafnium-free Half-Heuslers," *Materialia,* vol. 1, pp. 168-174, 2018.

[17] L. Zhao, Y. He, H. Zhang, L. Yi, and J. Wu, "Enhancing the thermoelectric property of Bi2Te3 through a facile design of interfacial phonon scattering," *Journal of Alloys and Compounds,* vol. 768, pp. 659-666, 2018.

[18] P. Cermak, P. Ruleova, V. Holy, J. Prokleska, V. Kucek, K. Palka, *et al.*, "Thermoelectric and magnetic properties of Cr-doped single crystal Bi2Se3 – Search for energy filtering," *Journal of Solid State Chemistry,* vol. 258, pp. 768-775, 2018.

[19] A. Pakdel, Q. Guo, V. Nicolosi, and T. Mori, "Enhanced thermoelectric performance of Bi–Sb–Te/Sb2O3 nanocomposites by energy filtering effect," *Journal of Materials Chemistry A,* vol. 6, pp. 21341-21349, 2018.

[20] J. Peng, L. Fu, Q. Liu, M. Liu, J. Yang, D. Hitchcock, *et al.*, "A study of Yb0.2Co4Sb12–AgSbTe2nanocomposites: simultaneous enhancement of all three thermoelectric properties," *J. Mater. Chem. A,* vol. 2, pp. 73-79, 2014.

[21] Q. Zhang, B. Liao, Y. Lan, K. Lukas, W. Liu, K. Esfarjani, *et al.*, "High thermoelectric performance by resonant dopant indium in nanostructured SnTe," *Proc Natl Acad Sci U S A,* vol. 110, pp. 13261-6, Aug 13 2013.

[22] M. Zebarjadi, K. Esfarjani, M. S. Dresselhaus, Z. F. Ren, and G. Chen, "Perspectives on thermoelectrics: from fundamentals to device applications," *Energy Environ. Sci.,* vol. 5, pp. 5147-5162, 2012.

[23] S. V. Faleev and F. Léonard, "Theory of enhancement of thermoelectric properties of materials with nanoinclusions," *Physical Review B,* vol. 77, 2008.

[24] A. Popescu, L. M. Woods, J. Martin, and G. S. Nolas, "Model of transport properties of thermoelectric nanocomposite materials," *Physical Review B,* vol. 79, 2009.

[25] G. E. Pike and C. H. Seager, "The dc voltage dependence of semiconductor grain‐boundary resistance," *Journal of Applied Physics,* vol. 50, pp. 3414-3422, 1979.

[26] D. M. Rowe and G. Min, "Multiple potential barriers as a possible mechanism to increase the Seebeck coefficient and electrical power factor," vol. 316, pp. 339-342, 1994.

[27] D. Narducci, S. Frabboni, and X. Zianni, "Silicon de novo: energy filtering and enhanced thermoelectric performances of nanocrystalline silicon and silicon alloys," *Journal of Materials Chemistry C,* vol. 3, pp. 12176-12185, 2015.

[28] A. Singha and B. Muralidharan, "Incoherent scattering can favorably influence energy filtering in nanostructured thermoelectrics," *Sci Rep,* vol. 7, p. 7879, Aug 11 2017.

[29] J.-H. Bahk, Z. Bian, and A. Shakouri, "Electron energy filtering by a nonplanar potential to enhance the thermoelectric power factor in bulk materials," *Physical Review B,* vol. 87, 2013.

[30] Y. Xia, J. Park, F. Zhou, and V. Ozoliņš, "High Thermoelectric Power Factor in Intermetallic CoSi Arising from Energy Filtering of Electrons by Phonon Scattering," *Physical Review Applied,* vol. 11, 2019.

[31] M. T. Dylla, J. J. Kuo, I. Witting, and G. J. Snyder, "Grain Boundary Engineering Nanostructured SrTiO3 for Thermoelectric Applications," *Advanced Materials Interfaces,* p. 1900222, 2019.

[32] J. J. Kuo, S. D. Kang, K. Imasato, H. Tamaki, S. Ohno, T. Kanno, *et al.*, "Grain boundary dominated charge transport in Mg3Sb2-based compounds," *Energy & Environmental Science,* vol. 11, pp. 429-434, 2018.

[33] J. Li, Q. Tan, J.-F. Li, D.-W. Liu, F. Li, Z.-Y. Li, *et al.*, "BiSbTe-Based Nanocomposites with HighZT: The Effect of SiC Nanodispersion on Thermoelectric Properties," *Advanced Functional Materials,* vol. 23, pp. 4317-4323, 2013.

[34] T. C. Harman, "Special Techniques for Measurement of Thermoelectric Properties," *Journal of Applied Physics,* vol. 29, p. 1373, 1958.

[35] "Sentaurus Device User Guide," Sentaurus Device simulation software, Synopsys Corp., 2019. https://www.synopsys.com/silicon/tcad/device-simulation/sentaurus-device.html

[36] G. K. Wachutka, "Rigorous thermodynamic treatment of heat generation and conduction in semiconductor device modeling," *IEEE Transactions on Computer-Aided Design of Integrated Circuits and Systems,* vol. 9, pp. 1141-1149, 1990.

[37] S. Shigetomi and S. Mori, "Electrical Properties of Bi2Te3," *Journal of the Physical Society of Japan,* vol. 11, pp. 915-919, 1956.

[38] P. Giannozzi, S. Baroni, N. Bonini, M. Calandra, R. Car, C. Cavazzoni, *et al.*, "QUANTUM ESPRESSO: a modular and open-source software project for quantum simulations of materials," *J Phys Condens Matter,* vol. 21, p. 395502, Sep 30 2009.

[39] X. Wang, V. Askarpour, J. Maassen, and M. Lundstrom, "On the calculation of Lorenz numbers for complex thermoelectric materials," *Journal of Applied Physics,* vol. 123, p. 055104, 2018.

[40] M. Winkler, X. Liu, J. D. König, S. Buller, U. Schürmann, L. Kienle, *et al.*, "Electrical and structural properties of Bi2Te3 and Sb2Te3 thin films grown by the nanoalloying method with different deposition patterns and compositions," *Journal of Materials Chemistry,* vol. 22, p. 11323, 2012.

[41] D.-H. Kim and T. Mitani, "Thermoelectric properties of fine-grained Bi2Te3 alloys," *Journal of Alloys and Compounds,* vol. 399, pp. 14-19, 2005.

[42] R. Ionescu, J. Jaklovszky, N. Nistor, and A. Chiculita, "Grain size effects on thermoelectrical properties of sintered solid solutions based on Bi2Te3," *Physica Status Solidi (a),* vol. 27, pp. 27-34, 1975.

[43] L.-D. Zhao, B.-P. Zhang, W.-S. Liu, and J.-F. Li, "Effect of mixed grain sizes on thermoelectric performance of Bi2Te3 compound," *Journal of Applied Physics,* vol. 105, p. 023704, 2009.

[44] O. Ivanov, O. Maradudina, and R. Lyubushkin, "Grain size effect on electrical resistivity of bulk nanograined Bi 2 Te 3 material," *Materials Characterization,* vol. 99, pp. 175-179, 2015.

[45] M. Lossec, Bernard Multon, Hamid Ben Ahmed, and C. Goupil, "Thermoelectric generator placed on the human body: system modeling and energy conversion improvements," *Eur. Phys. J. Appl. Phys.,* vol. 52, 2010.

[46] Morgan Hedden, Nick Francis, Jason T Haraldsen, T. Ahmed, and C. Constantin, "Thermoelectric Properties of Nano-Meso-Micro β-MnO2 Powders as a Function of Electrical Resistance," *Nanoscale Research Letters,* p. 292, 2015.